\RequirePackage{lineno}
\documentclass[twocolumn,showpacs,aps,prd,superscriptaddress,floatfix]{revtex4-1}
\usepackage{graphicx}  
\usepackage{dcolumn}   
\usepackage{bm}        
\usepackage{amssymb}   
\usepackage{amsfonts}  
\usepackage{hyperref}
\usepackage{amsmath}
\usepackage{mathtools}
\usepackage{xspace}
\usepackage{color}
\usepackage{mathrsfs}
\def \DzDzbar  {D^{0}\bar{D^{0}}}
\def \DDbar    {D\bar{D}}
\def \DpDm     {D^{+}D^{-}}
\def \Dp       {D^{+}}
\def \Dm       {D^{-}}
\def \Dz       {D^{0}}
\def \Dzbar    {\bar{D}^0}
\def \Dbar     {\bar{D}}
\def \Ks       {K_S^0}
\def \Kp       {K^+}
\def \Km       {K^-}
\def \Kpm      {K^\pm}
\def \pip      {\pi^+}
\def \pim      {\pi^-}
\def \piz      {\pi^0}
\def \pipm     {\pi^\pm}

\def \jpsi     {J/\psi}

\def \psipp    {\psi(3770)}
\def \eff      {\varepsilon}

\def \BR       {\mathcal{B}}

\def \Mbcsig   {M_{\rm BC}^{\rm sig}}
\def \Mbctag   {M_{\rm BC}^{\rm tag}}
\def \De       {\Delta E}

\def \Ebeam    {E_{\rm beam}}
\def \pppm     {\pi^+\pi^-}
\def \pppz     {\pi^+\pi^0}
\def \pmpz     {\pi^-\pi^0}
\def \pzpz     {\pi^0\pi^0}
\def \qqbar    {q\bar{q}}
\def \ppp      {\pi^+\pi^-\pi^0}
\def \wpipi    {\omega \pi\pi}
\def \wpppm    {\omega \pi^+\pi^-}
\def \wpppz    {\omega \pi^+\pi^0}

\def \wpzpz    {\omega \pi^0\pi^0}
\def \etapipi  {\eta \pi\pi}

\def \gev  {\mbox{GeV}}

\def \gevcc{\mbox{GeV/$c^2$}}
\def \mev  {\mbox{MeV}}

\def \mevcc{\mbox{MeV/$c^2$}}
\def \ifb  {\mbox{fb$^{-1}$}}

\def \dz         {D^{0}}
\def \dzbar   {\bar{D}^{0}}
\def \dbar   {\bar{D^{0}}}
\def \de      {\Delta E}

\def \wpipi    {\omega \pi\pi}

\def \ee   {e^+e^-}
\def \gev  {\mbox{GeV}}

\def \gevcc{\mbox{GeV/$c^2$}}
\def \mev  {\mbox{MeV}}
\def \mevcc{\mbox{MeV/$c^2$}}
\def \ifb  {\mbox{fb$^{-1}$}}

\def \epem {e^+e^-}

\def \pzpz {\pi^0\pi^0}
\def \pppm {\pi^{+}\pi^{-}}
\def \piz  {\pi^0}
\def \pip  {\pi^+}
\def \pim  {\pi^-}

\begin{document}
\title{\bf  Measurement of Singly Cabibbo-Suppressed Decays \boldmath{$D \to \wpipi$}}

\author{
		\begin{center}
	M.~Ablikim$^{1}$, M.~N.~Achasov$^{10,c}$, P.~Adlarson$^{64}$, S. ~Ahmed$^{15}$, M.~Albrecht$^{4}$, A.~Amoroso$^{63A,63C}$, Q.~An$^{60,48}$, ~Anita$^{21}$, Y.~Bai$^{47}$, O.~Bakina$^{29}$
% ...
	\\
		\vspace{0.2cm}
		(BESIII Collaboration)\\
		\vspace{0.2cm} {\it
	$^{1}$ Institute of High Energy Physics, Beijing 100049, People's Republic of China\\
% ...
$^{4}$ Bochum  Ruhr-University, D-44780 Bochum, Germany\\
% ...
$^{10}$ G.I. Budker Institute of Nuclear Physics SB RAS (BINP), Novosibirsk 630090, Russia\\
% ...
$^{15}$ Helmholtz Institute Mainz, Johann-Joachim-Becher-Weg 45, D-55099 Mainz, Germany\\
% ...
$^{21}$ Indian Institute of Technology Madras, Chennai 600036, India\\
% ...
$^{29}$ Joint Institute for Nuclear Research, 141980 Dubna, Moscow region, Russia\\
% ...
$^{47}$ Southeast University, Nanjing 211100, People's Republic of China\\$^{48}$ State Key Laboratory of Particle Detection and Electronics, Beijing 100049, Hefei 230026, People's Republic of China\\
% ...
$^{60}$ University of Science and Technology of China, Hefei 230026, People's Republic of China\\
% ...
$^{63}$ (A)University of Turin, I-10125, Turin, Italy; (B)University of Eastern Piedmont, I-15121, Alessandria, Italy; (C)INFN, I-10125, Turin, Italy\\$^{64}$ Uppsala University, Box 516, SE-75120 Uppsala, Sweden\\
% ...
$^{c}$ Also at the Novosibirsk State University, Novosibirsk, 630090, Russia\\
% ...
		}\end{center}
		\vspace{0.4cm}
}

\begin{abstract}
   Using 2.93~fb$^{-1}$ of $e^{+}e^{-}$ collision data taken at a
   center-of-mass energy of 3.773\,GeV by the BESIII detector at the
   BEPCII, we measure the branching fractions of the singly
   Cabibbo-suppressed decays $D \to \wpipi$ to be $\BR(\Dz \to \wpppm)
   = (1.33 \pm 0.16 \pm 0.12)\times 10^{-3}$ and $\BR(\Dp \to \wpppz)
   =(3.87 \pm 0.83 \pm 0.25)\times 10^{-3}$, where the first
   uncertainties are statistical and the second ones systematic.  The
   statistical significances are $12.9\sigma$ and $7.7 \sigma$,
   respectively.  The precision of $\BR(\Dz \to \wpppm)$ is improved
   by a factor of 2.1 over prior measurements, and $\BR(\Dp\to
   \wpppz)$ is measured for the first time.  No significant signal for
   $\Dz \to \wpzpz$ is observed, and the upper limit on the branching
   fraction is $\BR(\Dz \to \wpzpz) < 1.10 \times 10^{-3}$ at the 90\%
   confidence level. The branching fractions of $D\to \etapipi$ are
   also measured and consistent with existing results.
\end{abstract}


\maketitle
\section{Introduction}
The study of multi-body hadronic decays of charmed mesons is important
to understand the decay dynamics of both strong and weak
interactions. It also provides important input to the beauty sector
for the test of Standard Model (SM) predictions.  For instance, the
self-conjugate decay $\Dz \to \wpppm$ can be used to improve the
measurement of the Cabibbo-Kobayashi-Maskawa (CKM) angle $\gamma$ via
$B^{\pm} \to \dz \Kpm$~\cite{glw, ads, ggsz}, and the unmeasured decay
$\Dp \to \wpppz$ is a potential background in the semi-tauonic decay
$B \to D\tau \nu_{\tau}$. The ratio of branching fractions (BFs
) $\mathcal{R}(D)$, defined as
$\mathcal{B}(B\to D \tau \nu_{\tau})/\mathcal{B}(B \to D l
\nu_l)~(l=e,\mu)$, probes lepton flavor universality (LFU). The
current world average measurement of $\mathcal{R}(D)$ is around
$3.1\sigma$ away from the SM prediction~\cite{HFLAV,belle}, which is
evidence of LFU violation.  However, the BFs of many
multi-body hadronic decays, especially for singly Cabibbo-suppressed
(SCS) or doubly Cabibbo-suppressed (DCS) decays of $D$ mesons, are
still either unknown or imprecise due to low decay rates or huge
backgrounds.  Precise measurements of these decays are desirable in
several areas.

Until now, for the SCS decays $D\to\wpipi$, only the branching
fraction  of $\Dz \to \wpppm$ has been measured; CLEO found $B(\Dz
\to \wpppm) = (1.6 \pm 0.5)\times 10^{-3}$~\cite{Rubin}, where the
precision is limited by low statistics.  The data used in this
analysis is a $\psipp$ sample with an integrated luminosity of
2.93~$\ifb$~\cite{psipp} collected at a center-of-mass energy of
$3.773$~GeV with the BESIII detector at the BEPCII collider. It
provides an excellent opportunity to improve these measurements.
Furthermore, in the decay $\psipp \to \DzDzbar$, the $\Dz$ and
$\Dzbar$ mesons are coherent and of opposite $CP$ eigenvalues. Thus,
a sufficiently large sample can also be used to measure the fractional
$CP$-content of the decay $\Dz \to \wpppm$, which is necessary to
relate the $CP$-violating observables to the CKM angle $\gamma$ via
the so-called quasi-GLW method~\cite{glw}.

In this paper, we present measurements of the absolute BFs of the SCS
decays $D \to \wpipi$ with the \rm{\lq\lq double tag\rq\rq} (DT)
technique, pioneered by the MARK-III collaboration~\cite{mark3}. The advantage
of this technique is to reduce the combinatorial backgrounds from non-$D\bar{D}$ decays
with a cost of loss of the statistics.  
The $\omega$ mesons are reconstructed in the $\ppp$ final states. We also
measure the BFs for $D \to \etapipi$ with the subsequent decay $\eta
\to \ppp$, which are used to verify the results measured with the
$\eta \to \gamma \gamma$ decay mode and theoretical
models~\cite{Peshkin,Gronau}.
Throughout the paper, the charge conjugate modes are always implied, unless explicitly stated.

\section{BESIII Detector and Monte Carlo Simulation}
BESIII is a cylindrical spectrometer covering $93\%$ of
the total solid angle.  It consists of a helium-gas-based main drift
chamber (MDC), a plastic scintillator time-of-flight (TOF) system, a
CsI(Tl) electromagnetic calorimeter (EMC), a superconducting solenoid
providing a 1.0~T magnetic field, and a muon counter.  The momentum
resolution of a charged particle in the MDC is 0.5\% at a transverse
momentum of 1 GeV/$c$, and the energy resolution of a photon in the
EMC is 2.5(5.0)\% at 1~$\gev$ in the barrel (end-cap) region. Particle
identification (PID) is performed by combining the ionization energy
loss ($dE/dx$) measured by the MDC and the information from TOF.  The
details about the design and detector performance are provided in
Ref.~\cite{besdet}.

Monte Carlo (MC) simulation based on {\sc Geant4}~\cite{GEANT} is used
to optimize the event selection criteria, study the potential
backgrounds and evaluate the detection efficiencies.  The generator
{\sc KKMC}~\cite{2001SJadach} simulates the $\ee$ collision
incorporating the effects of beam energy spread and
initial-state-radiation (ISR).  An inclusive MC sample, containing
$\DDbar$ and non-$\DDbar$ events, ISR production
of $\psi(3686)$ and $\jpsi$, and continuum processes $\ee\to\qqbar$
($q= u, d, s$), is used to study the potential backgrounds.  The known
decays as specified in the Particle Data Group (PDG)~\cite{pdg} are
simulated by {\sc EvtGen}~\cite{evtgen}, while the remaining unknown
decays by {\sc LundCharm}~\cite{lundcharm}.

\section{Analysis Strategy}
We first select \rm{\lq\lq Single Tag\rq\rq} (ST) events in which the
$D$ meson candidate is reconstructed in a specific hadronic decay
mode.  Then the $D$ meson candidate of interest is reconstructed with
the remaining tracks. The absolute BFs for DT $D$ decays are calculated
by,
 \begin{eqnarray}
 \BR^{\rm sig} &=& \frac{N^{\rm sig}_{\rm DT}}{\BR^{\rm int} \, \sum\limits_i N^{i}_{\rm ST} ~ {\rm \eff}^{{\rm sig},i}_{\rm DT}~/~{\rm \eff}^{i}_{\rm ST}},
 \label{eq:absBr}
\end{eqnarray}
where $N^{\rm sig}_{\rm DT}$ and $N^{i}_{\rm ST}$ are the yields of
  DT signal events and ST events, $\eff^{i}_{\rm ST}$ and $\eff^{{\rm
      sig}, i}_{\rm DT}$ are the ST and DT detection efficiencies for
  a specific ST mode $i$, respectively, and $\BR^{\rm int}$ is the
  product of the BFs of the intermediate states $\omega/\eta$ and
  $\pi^0$ in the subsequent decays of the $D$ meson.

\section{Data Analysis}
For each tag mode, the $D$ meson candidates are reconstructed from all
possible combinations of final state particles with the following
selection criteria. Charged tracks, not utilized for $K_S^0$
reconstruction, are required to have their distance of closest
approach to the interaction point (IP) be within 1~cm in the plane
perpendicular to the beam and $\pm 10$~cm along the beam.  The polar
angle $\theta$ with respect to the z-axis is required to satisfy
$|\cos\theta|< 0.93$.  PID is performed to determine likelihood
$\mathcal{L}$ values for the $\pipm$ and $\Kpm$ hypotheses, and
$\mathcal{L}_{\pi} > \mathcal{L}_{K}$ and $\mathcal{L}_{K} >
\mathcal{L}_{\pi}$ are required for the $\pipm$ and $\Kpm$ candidates,
respectively.

The $\Ks$ candidates are reconstructed from a pair of oppositely charged
tracks.  These two tracks are assumed to be pions without performing
PID and are required to be within $\pm 20$~cm from the IP along the
beam direction, but with no constraint in the transverse plane.  A
fit of the two pions to a common vertex is performed, and a
$\Ks$ candidate is required to have a $\chi^2$ of the
vertex-constrained fit less than 100.  The $\pppm$ invariant mass
$M_{\pppm}$ is required to be within three standard deviations from
the $\Ks$ nominal mass~\cite{pdg}, $0.487 < M_{\pppm} <
0.511~\gevcc$. The decay length of each selected $\Ks$ candidate
should be further than two standard deviations from the IP.

Photon candidates are reconstructed from clusters of energy deposits
in the EMC. The energy deposited in nearby TOF counter is included to
improve the reconstruction efficiency and energy resolution.  The
energy of each photon is required to be larger than 25~$\mev$ in the
barrel region ($|\cos\theta| < 0.8$) or 50~$\mev$ in the end-cap
region ($0.86 < |\cos\theta| <0.92$).  The EMC timing of the photon is
required to be within 700 ns relative to the event start time to
suppress electronic noise and energy deposits unrelated to the event.
A $\piz$ candidate is reconstructed from a photon pair with an
invariant mass within $[0.115,~ 0.150]~\gevcc$, and at least one
photon should be detected in the EMC barrel region.  To improve the
momentum resolution, a kinematic fit is carried out constraining the
invariant mass of the selected photon pair to the $\pi^0$ nominal
mass~\cite{pdg}, and the resultant kinematic variables are used in the
subsequent analysis.

In this analysis, the ST events are selected by reconstructing
$\dzbar$ candidates with $\Kp\pim, \Kp\pmpz$ and $\Kp\pim\pim\pip$
final states and $D^-$ candidates with $\Kp \pim\pim$, $\Kp
\pim\pim\piz$, $\Ks \pim$, $\Ks \pim\piz$, $\Ks \pim \pim \pip$ and
$\Kp\Km\pim$ final states, which comprise approximately $26\%$ and
$28\%$ of total $\Dzbar$ and $\Dm$ (referred to as $\Dbar$ later)
decays, respectively.  Two variables, the energy difference $\De
\equiv E_{\rm D} - \Ebeam$ and the beam-constrained mass
$\Mbctag\equiv \sqrt{\Ebeam^{2}/c^{4} - p^{2}_{\rm D}/ c^{2}}$, are
used to identify the $\bar{D}$ candidates. Here $\Ebeam$ is the beam
energy, and $E_{\rm D}(p_{\rm D})$ is the reconstructed energy
(momentum) of the $\bar{D}$ candidate in the $\epem$ center-of-mass
system.  The successful $\bar{D}$ candidate must satisfy
$\Mbctag>1.84~\gevcc$ and a mode-dependent $\De$ requirement, which is
approximately three times its resolution.
For an individual ST mode, if there are multiple candidates in an
event, the one with the minimum $|\de|$ is selected.  In the decay
process $\Dzbar\to \Kp\pim\pppm$, to remove backgrounds from $\Dzbar
\to \Ks \Kp\pim$, the invariant mass of any $\pppm$ is required to
satisfy $|M_{\pi^+\pi^-}-M_{K_S^0}|>30$ \mevcc, where $M_{K_S^0}$ is
the nominal mass of $\Ks$ ~\cite{pdg}.

To determine the ST yield, a binned maximum likelihood fit is
performed to the $\Mbctag$ distribution of selected candidate events
for each ST mode. The signal is described by the MC simulated shape
convolved with a Gaussian function which accounts for the
resolution difference between data and MC simulation, and the
combinatorial background is described by an ARGUS
function~\cite{argus} with a fixed endpoint parameter $\Ebeam$.  The
fit curves are presented in Fig.~\ref{fig:stFit}.

\begin{figure}[htbp]
  \centering
  \includegraphics[width=0.5\textwidth]{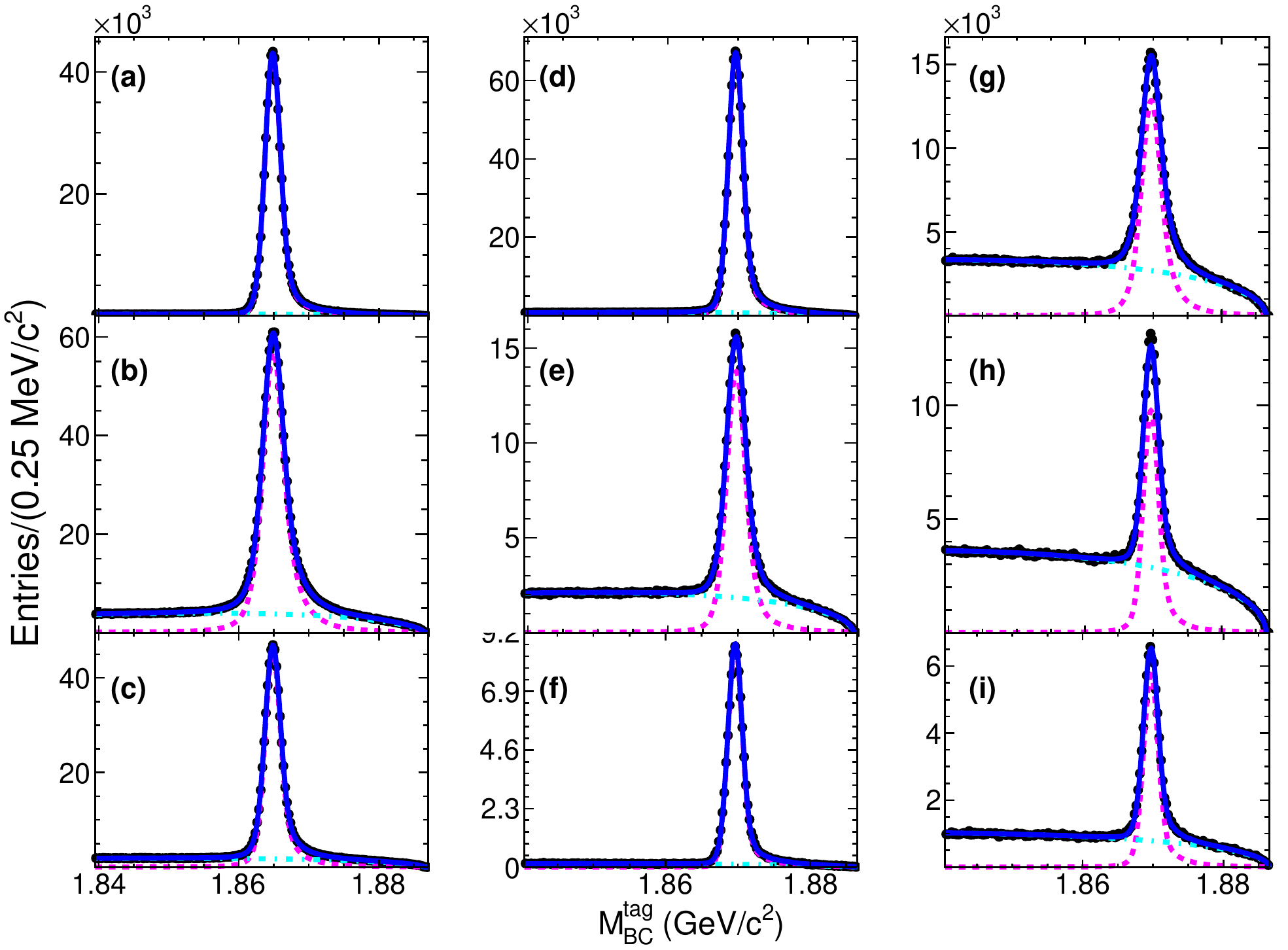}
  \caption{(color online) Fits to the $\Mbctag$ distributions for the
    ST modes: (a) $\Dzbar\to \Kp\pim$, (b) $\Dzbar\to \Kp\pmpz$, (c)
    $\Dzbar \to \Kp\pim\pim\pip$, (d) $\Dm \to \Kp \pim\pim$, (e) $\Dm
    \to \Kp\pim\pim\piz$, (f) $\Dm\to\Ks\pim$, (g) $\Dm \to \Ks
    \pim\piz$, (h) $\Dm \to \Ks\pim\pim\pip$ and (i)
    $\Dm\to\Kp\Km\pim$. Black dots with error bars represent data,
    green dashed-dot curves are the combinatorial background, red
    dashed curves are the signal shape and the blue solid curves are
    the total fit curves.}
  \label{fig:stFit}
\end{figure}

\begin{table*}[htbp]
  \begin{center}
  \footnotesize
  \caption{The ST yields in
    data ($N_{\rm ST}$), the efficiencies for ST ($\varepsilon_{\rm ST}$ in \%) and DT ($\varepsilon^{\rm modes}_{\rm DT}$ in \%) for $\dzbar$ decays. The uncertainties are statistical only. }
  \begin{tabular}{l c c c c c c}
    \hline \hline
    Mode                    &   $N^{i}_{\rm ST}$         & $\varepsilon^i_{\rm ST}$  & $\varepsilon_{\rm DT}^{\omega\pi^+\pi^-}$  & $\varepsilon_{\rm DT}^{\eta \pi^+\pi^-}$  & $\varepsilon_{\rm DT}^{\omega\pi^0\pi^0}$  & $\varepsilon_{\rm DT}^{\eta \pi^0\pi^0}$   \\ \hline
 $K^{+}\pi^{-}$               &  $542900 \pm 780$   & $66.7 \pm 0.02$    & $11.18 \pm 0.08$  & $12.94 \pm 0.07$  & $0.58 \pm 0.02$  & $0.56 \pm 0.02$              \\
 $K^{+}\pi^{-}\piz$           & $1065800 \pm 1280$  & $35.1 \pm 0.01$    & $5.92 \pm 0.06$    & $6.73  \pm 0.02$  & $0.31 \pm 0.01$  & $0.25 \pm 0.01$               \\
  $K^{+}\pi^{-}\pi^-\pi^+$    & $606310 \pm 890$    & $33.5 \pm 0.01$    & $4.48 \pm 0.06$    &  $5.08 \pm 0.04$  & $0.22 \pm 0.01$  & $0.23  \pm 0.01$              \\
     \hline\hline
  \end{tabular}
  \label{tab:stYield_Eff1}
  \end{center}
\end{table*}

\begin{table}[htbp]
  \begin{center}
  \footnotesize
  \caption{ The ST yields in data ($N_{\rm ST}$), the efficiencies for
    ST ($\varepsilon_{\rm ST}$ in \%) and DT ($\varepsilon^{\rm modes}_{\rm
      DT}$ in \%) for $D^-$ decays. The uncertainties are statistical
    only. }
  \begin{tabular}{l c c c c }
    \hline \hline
    Mode                    &   $N^{i}_{\rm ST}$         & $\varepsilon^i_{\rm ST}$   & $\varepsilon_{\rm DT}^{\omega\pi^+\pi^0}$  & $\varepsilon_{\rm DT}^{\eta \pi^+\pi^0}$  \\ \hline
 $K^+\pi^-\pi^-$            & $794890 \pm 959$    & $49.7 \pm 0.03$     & $2.47 \pm 0.06$    & $2.57 \pm 0.02$   \\
 $K^+\pi^-\pi^-\pi^0$       & $216720 \pm 609$    & $22.4 \pm 0.03$     & $0.56 \pm 0.04$    & $0.99 \pm 0.03$   \\
 $K_S^0\pi^-$               & $97769 \pm 333$     & $52.8 \pm 0.10$      & $2.30 \pm 0.06$    & $2.67 \pm 0.03$   \\
 $K_S^0\pi^-\pi^0$          & $224880 \pm 661$    & $27.6 \pm 0.04$      & $1.28 \pm 0.04$    & $1.29 \pm 0.04$   \\
 $K_S^0\pi^-\pi^-\pi^+$     & $130300 \pm 513$    & $36.0 \pm 0.07$     & $1.50 \pm 0.04$    & $1.56 \pm 0.04$   \\
 $K^-K^+\pi^-$              & $70299 \pm 326$     & $40.7 \pm 0.11$        & $2.39 \pm 0.06$    & $2.35 \pm 0.04$  \\
    \hline\hline
  \end{tabular}
  \label{tab:stYield_Eff2}
  \end{center}
\end{table}

The same procedure is used on the inclusive MC sample to determine
the ST efficiency. The corresponding ST yields and efficiencies for each
individual tag mode are summarized in Tables~\ref{tab:stYield_Eff1}
and ~\ref{tab:stYield_Eff2} for $\dzbar$ and $D^-$ decays,
respectively.  Here the yields for $\bar{D}^0 \to K^+\pi^-$,
$K^+\pi^-\pi^0$ and $K^+\pi^-\pi^+\pi^-$ decays include the
contributions from the DCS decays $\bar{D}^0 \to K^-\pi^+$,
$K^-\pi^+\pi^0$ and $K^-\pi^+\pi^-\pi^+$, respectively.

For the DT candidates, we further reconstruct the decays $\Dz \to \ppp
\pppm$ and $\ppp\pzpz$ as well as $\Dp \to \ppp \pppz$ using the
remaining $\pipm$ and $\piz$ candidates.  The corresponding $\De$ and
$\Mbcsig$ requirements distinguish signal candidates from
combinatorial backgrounds. The $\De$ distribution is required to be within 3.0 (3.5)  
times of its resolution for $\Dz \to \ppp \pppm$ ($\Dz \to  \ppp\pzpz$ and $\Dp \to \ppp \pppz$) decays.  For a given signal mode, if there are multiple combinations
in an event, the one with the minimum $|\De|$ is selected.  Since the
 signal final states contain multiple pions, an irreducible background
with the same final state is that from the Cabibbo-favored (CF)
processes including $\Ks \to \pi\pi$, and a candidate is vetoed if the
invariant mass of any $\pi\pi$ combination lies within the $\Ks$ mass
window, {\it i.e.}, $0.475 < M_{\pppm} < 0.520$ or $0.448 < M_{\pzpz}
< 0.548~\gevcc$.  Four possible $\ppp$ combinations exist in the
decays $\Dz\to \ppp \pppm$ and $\Dp \to \ppp \pppz$, while there are
three $\ppp$ combinations in $\Dz \to \ppp \pzpz$.  Combinations with
the invariant mass $M_{\pi^+\pi^-\pi^0}$ less than 0.9~$\gevcc$ are
retained for further analysis.  The inclusion of multiple combinations
for an event avoids peaking background in the $M_{\pi^+\pi^-\pi^0}$
distribution with a cost of additional combinatorial backgrounds.

\begin{figure}[!htbp]
  \centering
  \includegraphics[width=0.45\textwidth]{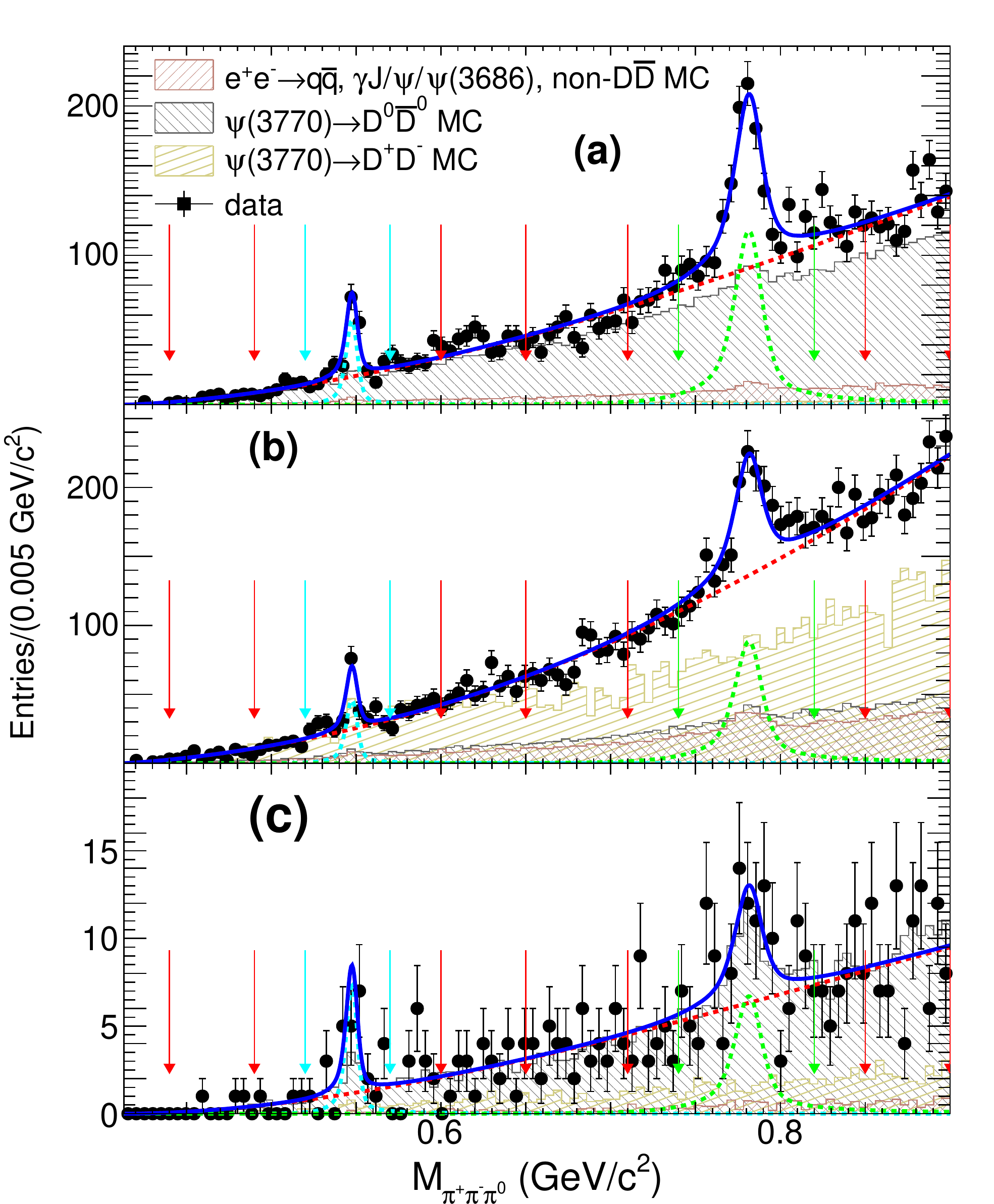}

  \caption{ Fits to the $M_{\ppp}$ distributions for the processes:
    (a) $\Dz\to \ppp\pppm$, (b) $\Dp \to \ppp \pppz$ and (c) $\Dz \to
    \ppp\pzpz$, together with the  background predictions from various MC samples shown by
    the histograms with  diagonal pattern lines. The MC samples of
    $\psi(3770)\to D\bar{D}$ decays include various background processes only.
    Black dots with error bars are data, dashed red curves
    are combinatorial background, dotted cyan and green curves are
    $\eta$ and $\omega$ signals, and the solid blue curves are the
    total fit curves. The two green and cyan arrow lines represent the
    $\omega$ and $\eta$ signal regions, respectively, and the two red
    arrows represent their low- and high-sideband regions.  }
 \label{fig:dtFit}
\end{figure}

After applying the above selection criteria in both ST and DT sides including $M_{\rm BC}^{\rm sig} > 1.84$ \gevcc, the $M_{\pi^+\pi^-\pi^0}$
distributions are shown in Fig.~\ref{fig:dtFit}, where the $\omega$
and $\eta$ signals are clear that might originate from either $D\to \omega/\eta \pi\pi$ decays or various background processes. The two-dimensional (2D) distribution
of $M_{\rm BC}^{\rm tag}$ versus $M_{\rm BC}^{\rm sig}$ is shown in
Fig.~\ref{2DHist}.  The signal of $\psi(3770)\to \DDbar$ (including
the background with the same final states, but without $\omega/\eta$
signals) is expected to concentrate around the intersection of $M_{\rm
  BC}^{\rm tag} = M_{\rm BC}^{\rm sig}=M_D$, where $M_D$ is the $D$
nominal mass.  The background events from $\psi(3770)\to \DDbar$ with
a correctly reconstructed $D$ meson and an incorrectly reconstructed
$\bar{D}$ meson (namely BKGI) distribute along the horizontal and
vertical bands with $M_{\rm BC}^{\rm tag}(M_{\rm BC}^{\rm sig})=M_D$.
The background events from the $e^+e^- \to q\bar{q}$ process (BKGII)
spread along the diagonal, and do not peak in either the $M_{\rm BC}^{\rm
  tag}$ or $M_{\rm BC}^{\rm sig}$ distribution.  A small background
including both $e^+e^- \to q\bar{q}$ and $\psi(3770)\to \DDbar$, with
neither $D$ nor $\Dbar$ correctly reconstructed (BKGIII), is assumed
to distribute uniformly in the $M_{\rm BC}^{\rm tag}$ versus $M_{\rm
  BC}^{\rm sig}$ phase space (PHSP).

 \begin{figure}[!htbp]
  \centering
  \includegraphics[width=0.45\textwidth]{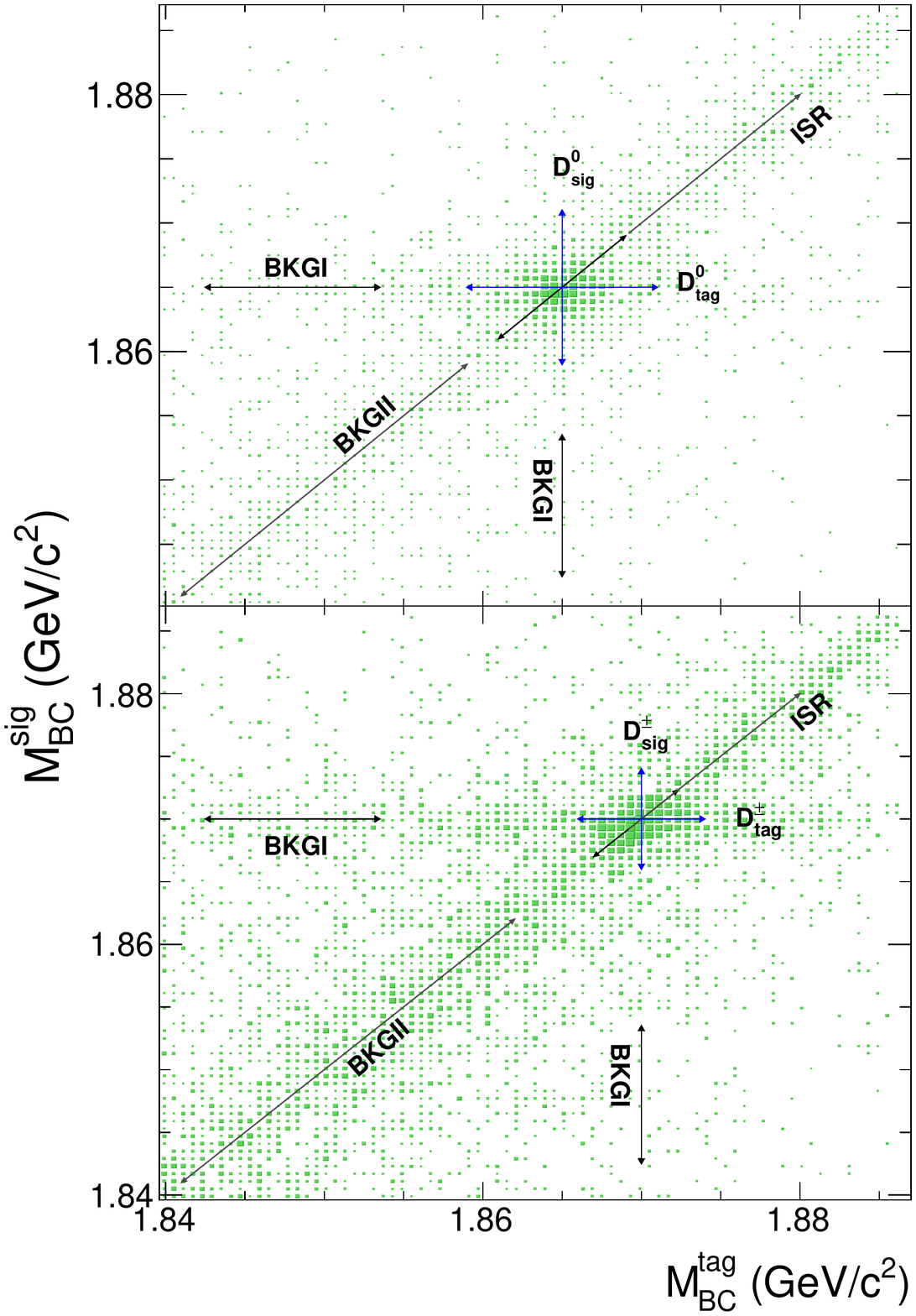}

  \caption{The 2D distributions of $M_{\rm BC}^{\rm tag}$ versus $M_{\rm BC}^{\rm sig}$ for the
   DT candidate events of $\psi(3770)\to \DzDzbar$ (top) and $\DpDm$ (bottom).}
 \label{2DHist}
\end{figure}

To determine the signal yields (including the background with same
final states but without $\omega/\eta$ signals), a 2D unbinned maximum
likelihood fit is performed to the $M_{\rm BC}^{\rm tag}$ versus
$M_{\rm BC}^{\rm sig}$ distribution of candidate events within the
$\omega(\eta)$ signal region, defined as $0.74(0.52)<
M_{\pi^+\pi^-\pi^0} < 0.82(0.57)$~\gevcc. The probability density
function (PDF) includes those of signal and three kinds of
backgrounds, described as:
\begin{widetext}
\begin{itemize}
\item Signal: $\mathscr{A}(M_{\rm BC}^{\rm sig},M_{\rm BC}^{\rm tag})$,
\item BKGI: $\mathscr{B}(M_{\rm BC}^{\rm tag})\times \mathscr{C}(M_{\rm BC}^{\rm sig}; E_{\rm beam},\xi_{M_{\rm BC}^{\rm sig}},\rho) + \mathscr{B}(M_{\rm BC}^{\rm sig}) \times \mathscr{C}(M_{\rm BC}^{\rm tag}; E_{\rm beam},\xi_{M_{\rm BC}^{\rm tag}},\rho)$,
\item BKGII: $\mathscr{C}((M_{\rm BC}^{\rm sig}+M_{\rm BC}^{\rm tag}); 2\cdot E_{\rm beam}, \xi,\rho)(\mathscr{F} \cdot G((M_{\rm BC}^{\rm sig}-M_{\rm BC}^{\rm tag});0,\sigma_{0}))+(1-\mathscr{F}) \cdot G((M_{\rm BC}^{\rm sig}-M_{\rm BC}^{\rm tag});0,\sigma_{1}))$,
\item BKGIII: $\mathscr{C}(M_{\rm BC}^{\rm sig}; E_{\rm beam},\xi_{M_{\rm BC}^{\rm sig}},\rho)\times \mathscr{C}(M_{\rm BC}^{\rm tag}; E_{\rm beam},\xi_{M_{\rm BC}^{\rm tag}},\rho)$,
\end{itemize}
\end{widetext}
 \noindent where $\mathscr{A}$
and $\mathscr{B}$ are 2D and one-dimensional (1D) signal PDFs for
$M_{\rm BC}^{\rm sig/tag}$ distributions, which are described with the
simulated signal shapes convolved with 2D and 1D Gaussian functions,
respectively, to account for the resolution difference between data and
MC simulation. $\mathscr{C}(x, E_{\rm end}, \xi,\rho)$ is an ARGUS
function~\cite{argus} with a fixed endpoint of $E_{\rm beam}$ and two
free parameters of $\xi$ and $\rho$. $\mathscr{F}$ is the fraction of
a Gaussian function $G(x;0,\sigma_i)$, the mean of which is zero and
the width $\sigma_{i}$ is $(M_{\rm BC}^{\rm sig}+M_{\rm BC}^{\rm
  tag})$ dependent: $\sigma_{i}=a_i(M_{\rm BC}^{\rm sig}+M_{\rm
  BC}^{\rm tag})+c_i$ ($i=0,1$). $\mathscr{F}$, $a_i$ and $c_i$ are
floated in the fit.  The projection plots of $M_{\rm BC}^{\rm
  tag}$ and $M_{\rm BC}^{\rm sig}$ are shown in
Fig.~\ref{fig:DTFit}, and the signal yields ($N^{\omega/\eta}_{\rm SG}$)
are summarized in Table~\ref{DTYield}.

\begin{figure*}[htbp]
  \centering
  \includegraphics[width=0.95\textwidth]{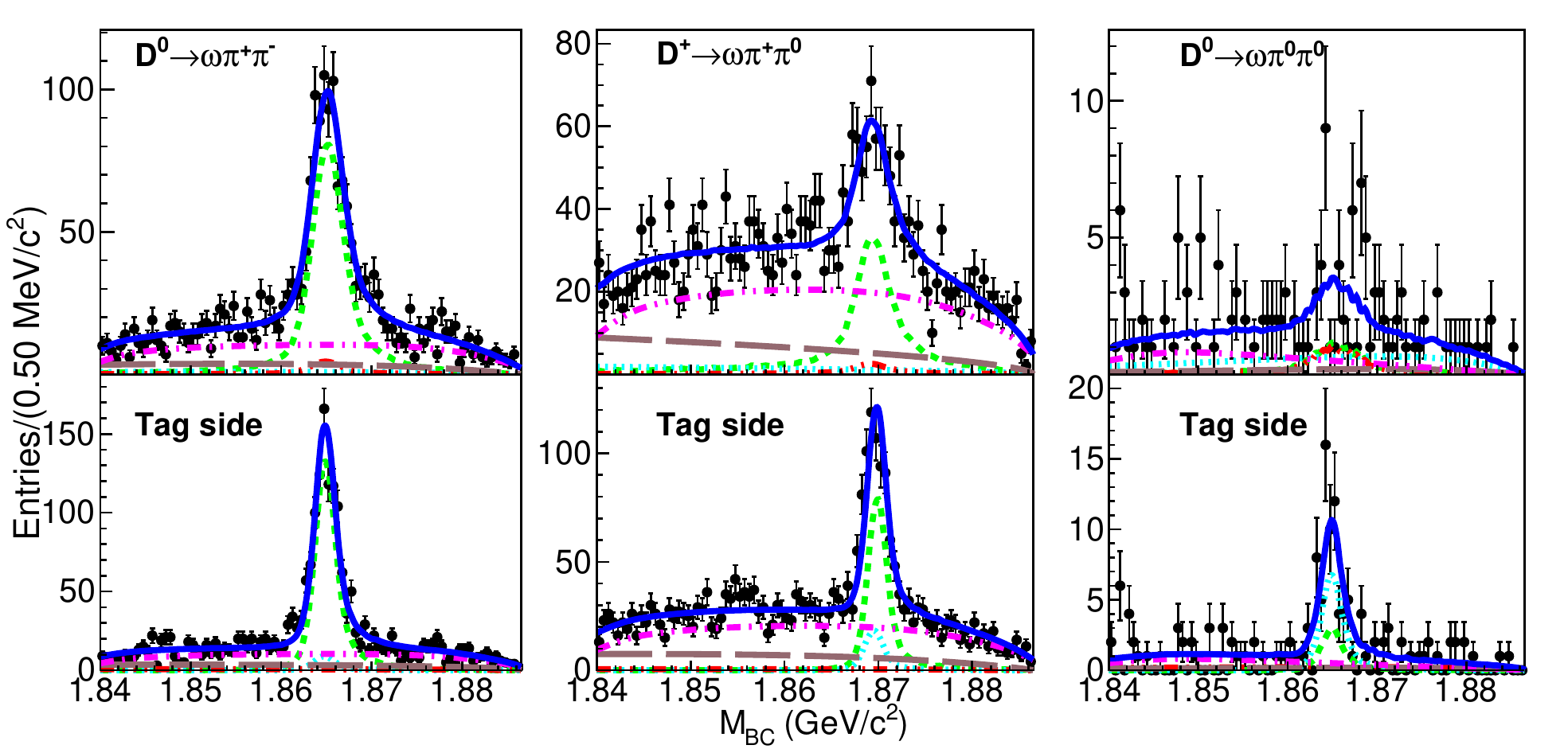}
 \includegraphics[width=0.95\textwidth]{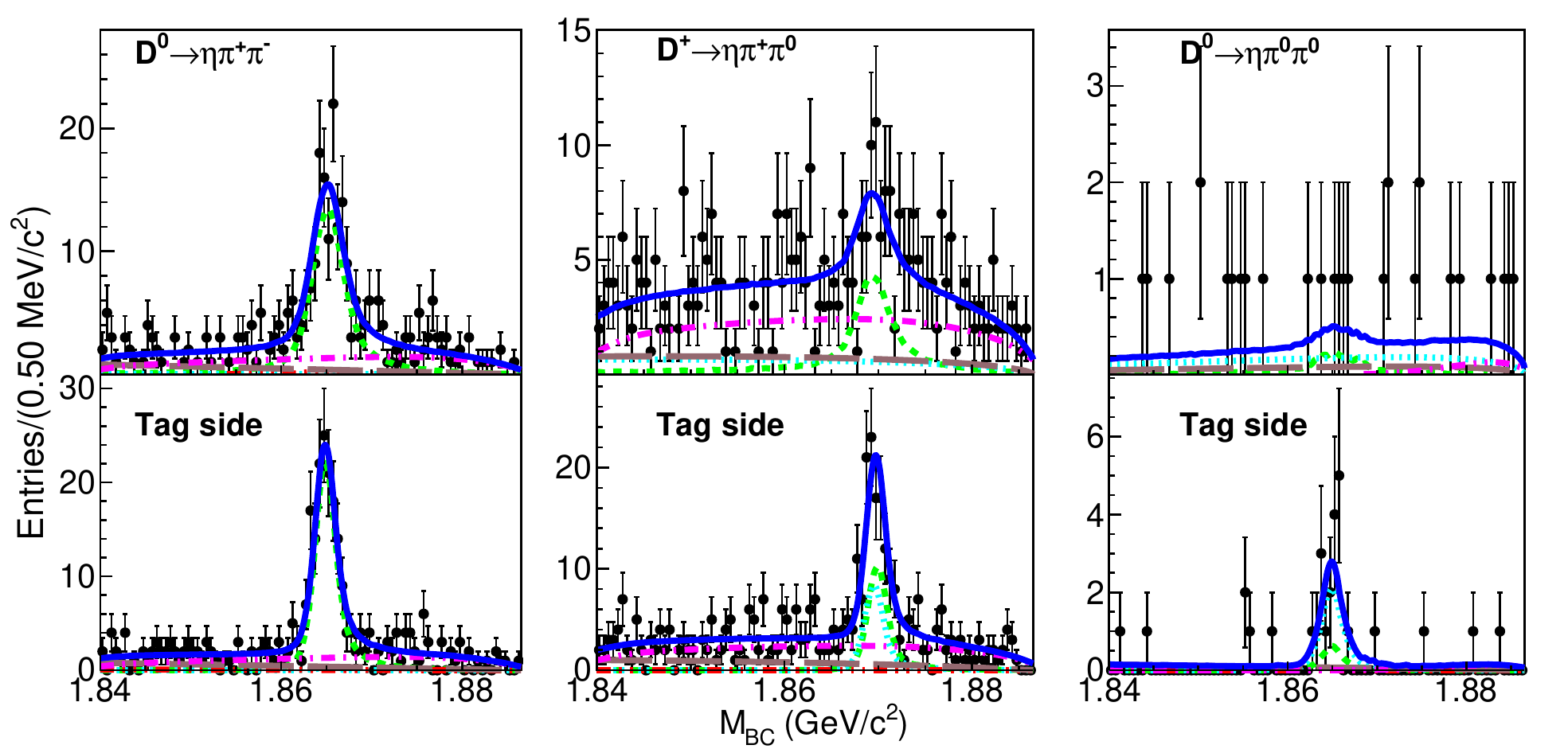}
    \caption{ Projection plots of the 2D fit to the distribution of
      $M_{\rm BC}^{\rm tag}$ versus $M_{\rm BC}^{\rm sig}$ for the DT
      candidate events in (top) $\omega$ and (bottom) $\eta$ signal
      regions. Black dots with error bars are data, the solid blue,
      dashed green, dotted cyan and dashed-dotted red, long
      dashed-dotted pink and long dashed brown curves represent the
      overall fit results, signal, BKGI, BKGII, and BKGIII,
      respectively. In each panel, the top plot is for $M_{\rm
        BC}^{\rm sig}$ and bottom for $M_{\rm BC}^{\rm tag}$. }
  \label{fig:DTFit}
\end{figure*}

To estimate the background with the same final states, but without
$\omega/\eta$ signal included (BKGIV), the same fit is performed on
the candidate events within the $\omega$ and $\eta$ sideband regions,
defined as $(0.65 < M_{\pi^+\pi^-\pi^0} < 0.71) \cup(0.85 <
M_{\pi^+\pi^-\pi^0} < 0.90)$~$\gevcc$ and $(0.44 < M_{\pi^+\pi^-\pi^0}
< 0.49) \cup(0.60 < M_{\pi^+\pi^-\pi^0} < 0.65)$~$\gevcc$,
respectively. The corresponding fit curves and signal yields
($N^{\omega/\eta}_{\rm SB}$) are shown in Fig.~\ref{fig:DTFit2} and
Table~\ref{DTYield}.  Additionally, there is also a small peaking
background from the CF processes $D^0 \to K_S^0 \omega/\eta$ (BKGV)
from events surviving the $K_S^0$ mass window veto due to its large
decay BF. The corresponding contributions ($N_{\rm peak}^{\rm BKGV}$)
are estimated by:
\begin{equation}
N_{\rm peak}^{\rm BKGV} = \mathcal{B} \cdot \sum\limits_i \frac{N^{i}_{\rm ST}  {\rm \varepsilon}^{i}_{\rm DT}}{{\rm \varepsilon}^{i}_{\rm ST}},
\label{BKGV}
\end{equation}
where $N^{i}_{\rm ST}$ and ${\rm \varepsilon}^{i}_{\rm ST}$ are the ST
yield and efficiency for tag mode $i$, respectively, as described in Eq.~\ref{eq:absBr}, $\mathcal{B}$ is
the product of the BFs of the decay $D^0 \to K_S^0 \omega/\eta$ as
well as its subsequent decays, taken from the PDG~\cite{pdg},
$\varepsilon^{i}_{\rm DT}$ is the DT detection efficiency for the $D^0
\to K_S^0 \omega/\eta$ decay, evaluated from exclusive MC samples.
The resultant $N_{\rm peak}^{\rm BKGV}$ for each individual process is
summarized in Table~\ref{DTYield}, where the uncertainties include
those from the BFs and statistics of the MC samples.

\begin{figure*}[htbp]
  \centering
   \includegraphics[width=0.95\textwidth]{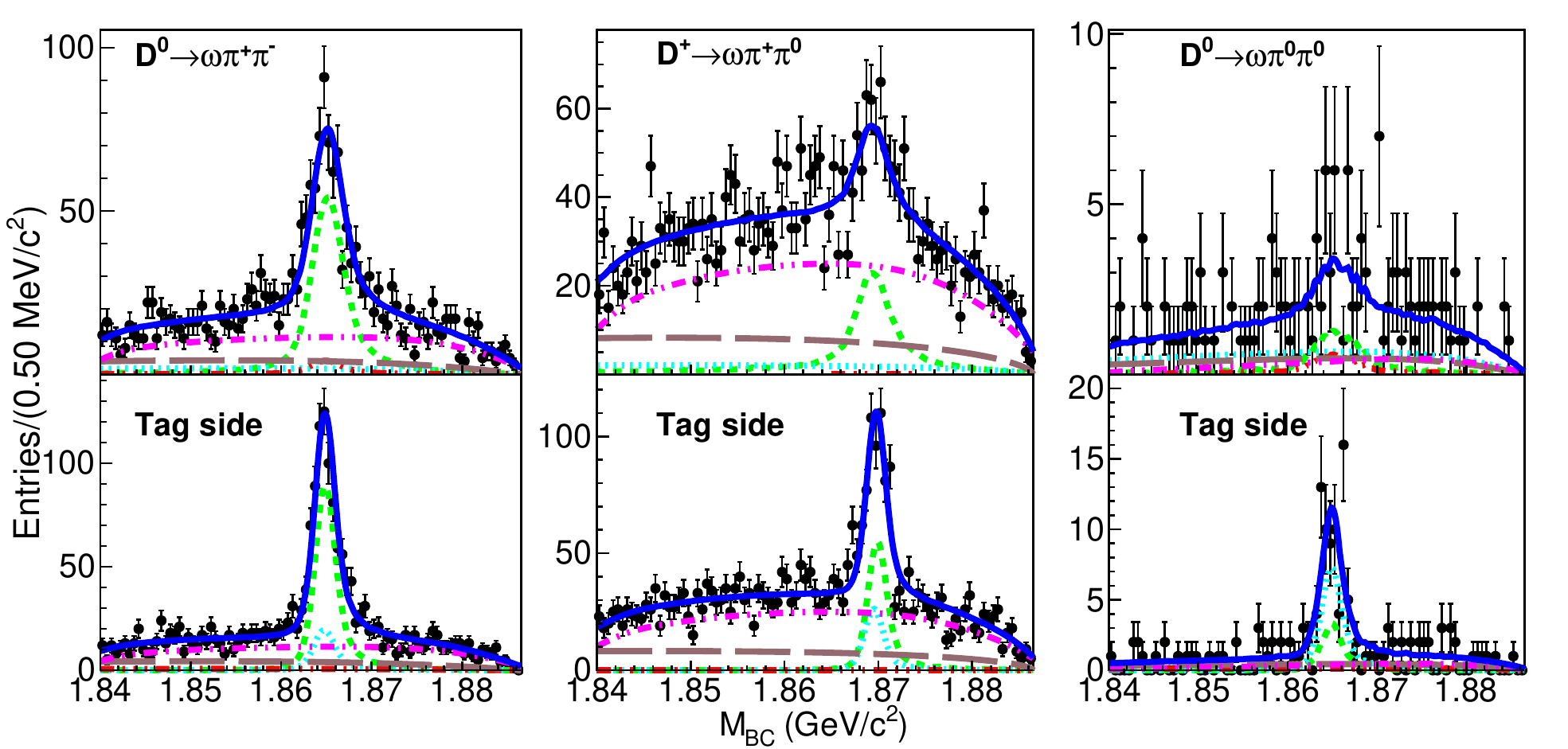}
    \includegraphics[width=0.95\textwidth]{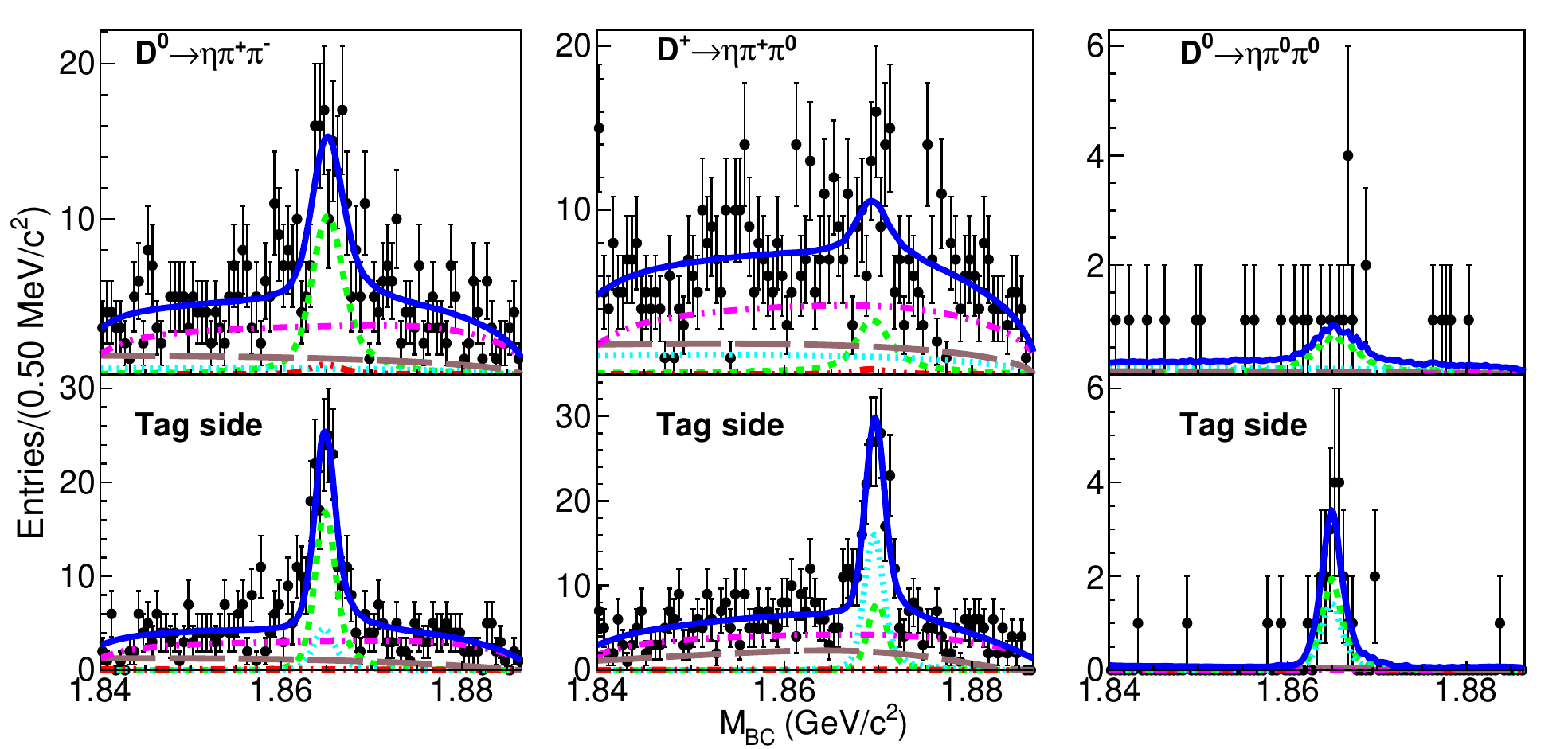}
  \caption{ Projection plots of the 2D fit to the distribution of
    $M_{\rm BC}^{\rm tag}$ versus $M_{\rm BC}^{\rm sig}$ for the DT
    candidate events in (top) $\omega$ and (bottom) $\eta$ sideband
    regions. Black dots with error bars are data, the solid blue,
    dashed green, dotted cyan and dashed-dotted red, long
    dashed-dotted pink and long dashed brown curves are the
    overall fit results, signal, BKGI, BKGII, and BKGIII,
    respectively. In each panel, the top plot is for $M_{\rm BC}^{\rm
      sig}$ and bottom for $M_{\rm BC}^{\rm tag}$.}
  \label{fig:DTFit2}
\end{figure*}

\begin{table*}[htbp]
  \begin{center} \footnotesize \caption{The yields of signal and
 individual backgrounds (see text) as well as the correction factor
 $f$, statistical significance (Sig.), $\BR^{\rm int}$ and BFs from
 this measurement and the PDG~\cite{pdg}.  Here and below, the
 first and second uncertainties are statistical and systematic,
 respectively. The upper limits are set at the $90\%$ C.L..}
\begin{tabular}{l c c c c c c c c c}\hline
 Decay mode                   & $N_{\rm SG}^{\omega/\eta}$ & $f~(\%)$ & $N_{\rm SB}^{\omega/\eta}$ & $N_{\rm peak}^{\rm BKGV}$ & $N_{\rm DT}^{\rm sig}$  & Sig.  & $\BR^{\rm int}$ & $\mathcal{B}^{\rm sig}~ (\times10^{-3})$   & $\mathcal{B}_{\rm PDG}~ (\times10^{-3})$ \\\hline
 $D^0 \to \omega \pi^+\pi^-$  &  $908.0 \pm 39.4$      & $74.6 \pm 1.5$   & $610.5 \pm 35.1$    & $41.4 \pm 2.5$   & $411.2 \pm 48.3$ & $12.9 \sigma$            & 0.882       & $1.33 \pm 0.16 \pm 0.12$  & $1.6 \pm 0.5$           \\
 $D^+ \to \omega \pi^+\pi^0$  &  $474.0 \pm 42.8$      & $73.3 \pm 1.2$   & $329.0 \pm 34.3$    & ---              & $232.9 \pm 49.8$ & $7.7 \sigma$                      &  0.872     & $3.87 \pm 0.83 \pm 0.25$  & ---   \\
 $D^0 \to \omega \pi^0\pi^0$  &  $20.2 \pm  10.5$      & $75.2 \pm 5.6$   & $22.1 \pm  10.0$    & $19.0 \pm 1.2$   & $-15.4  \pm 13.0$& $0.6 \sigma$                &  0.862   & $< 1.10$                        & --- \\     \hline

 $D^0 \to \eta \pi^+\pi^-$    & $151.3 \pm  14.6$      & $42.6 \pm 0.9$   & $115.0 \pm 15.3$    & $6.1 \pm 0.2$    & $96.2  \pm 16.0$  &$8.3\sigma$                     &   0.227   & $1.06 \pm 0.18 \pm 0.07$  & $1.09 \pm 0.16$ \\
 $D^+ \to \eta \pi^+\pi^0$    & $61.5 \pm   14.3$      & $41.4 \pm 0.7$   & $47.3 \pm  16.4$    & ---              & $41.9  \pm 15.8$  &$3.5\sigma$                              &   0.224  & $2.47 \pm 0.93 \pm 0.16$ & $1.38 \pm 0.35$ \\
 $D^0 \to \eta \pi^0\pi^0$    & $5.7 \pm 3.8$          & $40.6 \pm 3.3$   & $13.1 \pm  4.8$     & $2.0 \pm 0.1$    & $-1.6  \pm 4.3$   &$0.1 \sigma$                          &   0.221 & $< 2.38$  & $0.38 \pm  0.13$ \\  \hline
\end{tabular}
\label{DTYield}
\end{center}
 \end{table*}

The signal yield $N_{\rm DT}^{\rm sig}$ is given by
\begin{equation}
N_{\rm DT}^{\rm sig} = N_{\rm SG}^{\omega/\eta}-f\cdot N_{\rm SB}^{\omega/\eta}-N_{\rm peak}^{\rm BKGV},
\end{equation}
\noindent
where the correction factor $f$ is the ratio of background BKGIV yield
in the $\omega/\eta$ signal region to that in the sideband regions.
In practice, $f$ is determined by performing a fit to the
$M_{\pi^+\pi^-\pi^0}$ distribution, as shown in Fig.~\ref{fig:dtFit}.
In the fit, the $\omega/\eta$ signal is described by the sum of two
Crystal Ball functions~\cite{CB}, which have the same mean and
resolution values, but opposite side tails, and the background by a
reversed ARGUS function defined as Eq.~4 in Ref.~\cite{invisible} with
a fixed endpoint parameter corresponding to the $M_{\pi^+\pi^-\pi^0}$
threshold.  The signal DT efficiencies, as summarized in
Tables~\ref{tab:stYield_Eff1} and ~\ref{tab:stYield_Eff2} for $D^0$
and $D^+$ decays, respectively, are determined by the same approach on
the inclusive MC sample, which is the mixture of signal MC samples
generated with a unified PHSP distribution and various backgrounds.
Based on the above results, the decay BFs are calculated according to
Eq.~\ref{eq:absBr}, and are summarized in Table~\ref{DTYield}.  To
determine the statistical significance of signals for each individual
process, analogous fits are performed by fixing the signal yields to
those of the sum of backgrounds BKGIV and BKGV, and the resultant
likelihood values $\mathcal{L}_{0}$ are used to calculate the
statistical significance $\mathcal{S}=\sqrt{-2
  \ln(\mathcal{L}_0/\mathcal{L}_{\rm max})}$, as summarized in
Table~\ref{DTYield}, where $\mathcal{L}_{\rm max}$ is the likelihood
value of the nominal fit.

\section{Systematic Uncertainties}

According to Eq.~\ref{eq:absBr}, the uncertainties in the BF measurements include those associated with the detection efficiencies, ST and DT event yields as well as the BFs of the intermediate state decays.

With the DT method, the uncertainties associated with the detection
efficiency from the ST side cancel.  The uncertainty from the
detection efficiency of the signal side includes tracking, PID, $\piz$
reconstruction, $\Delta E$ requirement, $K_S^0$ veto and $\omega/\eta$
mass window requirement as well as the signal MC modeling.  The
uncertainties from the tracking, PID and $\piz$ reconstruction are
0.5\%, 0.5\%, and 2.0\%, respectively, which are obtained
by studying a DT control sample $\psi(3770) \to D\bar{D}$ with
hadronic decays of $D$ via a partial reconstruction
method~\cite{PID,Piz}.  The uncertainties associated with the $\Delta
E$ requirement, $K_S^0$ veto and $\omega/\eta$ mass window requirement
are studied with control samples of $D^0 \to 2(\pi^+\pi^-)\pi^0$,
$\pi^+\pi^-3\pi^0$ and $D^+ \to 2(\pi^+\pi^0)\pi^-$, which have the
same final state as the signal channels, include all possible
intermediate resonances and have higher yields than the signal
processes.  These control samples are selected with the DT method, and
their yields are obtained by fitting $M_{\rm BC}^{\rm sig}$
distributions.  To study the uncertainty from $\Delta E$, the control
samples are alternatively selected with a relatively loose $\Delta E$
requirement, $i.e.$ $|\Delta E|<0.1~\gev$, and then with the
nominal $\Delta E$ requirement. The ratio of the two signal yields
is taken as the corresponding efficiency. The same approach is
implemented with both data and the inclusive MC sample, and the
difference in efficiencies is taken as the uncertainty.  For the
$K_S^0$ veto uncertainty studies, we enlarge the $K_S^0$ veto mass
window of the control samples by 10~$\mevcc$, and the relative
difference in the efficiencies between data and inclusive MC sample is
taken as the uncertainty.  The uncertainties from the $\omega / \eta$
mass window requirement are studied by enlarging the corresponding
mass windows by 2~$\mevcc$ and the resulting difference in efficiency
between data and MC simulation is taken as the uncertainty.  In the
analysis, the three-body signal processes are simulated with the
uniform PHSP distribution, the corresponding uncertainties are
estimated with alternative MC samples, which assume $\pi\pi$ from
the $\rho$ resonance decay, and the resultant changes in efficiencies
are considered as the uncertainties.

The uncertainty related to the ST yield comes from the fit procedure,
and includes the signal and background shapes and the fit range.  The
uncertainty from the signal shape are estimated by alternatively
describing the signal with a kernel estimation~\cite{kernel} of the
signal MC derived shape convolved with a bifurcated Gaussian function.
The uncertainty from the background shape is estimated by
alternatively describing the shape with a modified ARGUS
function~\cite{argus} $(x^2/E_{\rm beam})(1-\frac{x^2}{E_{\rm
beam}^2})^{\rho}\cdot e^{\xi(1-\frac{x^2}{E_{\rm beam}^2})}$.  The
uncertainty from the fit range of $M^{\rm tag}_{\rm BC}$ is obtained
with a wider fit range, (1.835, 1.8865)~\gevcc.  The alternative fits
with the above different scenarios are performed, and the
resulting changes of signal yields are taken as the systematic
uncertainties.  The total uncertainties associated with the ST yields
are the quadrature sum of individual values.

\begin{table*}
  \begin{center}
  \footnotesize
  \caption{Systematic uncertainties and their sources. Here \rm{\lq Negl.\rq} means \rm{\lq Negligible\rq}.}
  \begin{tabular}{l |c c | c c | c c}
    \hline \hline
 Source   & \multicolumn{2}{|c|}{$D^0 \to \omega/\eta \pi^+\pi^-$} &  \multicolumn{2}{|c|}{$D^0 \to \omega/\eta \pi^0\pi^0$} &  \multicolumn{2}{|c}{ $D^+ \to \omega/\eta \pi^+\pi^0$} \\
 \hline 
  & ~~~$\omega\pi^+\pi^-$~~~ & ~~~$\eta \pi^+\pi^-$~~~  & ~~~$\omega \pi^0\pi^0$~~~ & ~~~$\eta \pi^0\pi^0$~~~ & ~~~$\omega \pi^+\pi^0$~~~ & ~~~$\eta \pi^+\pi^0$~~~ \\ \hline
\multicolumn{7}{c}{Additive systematic uncertainties (events) } \\
\hline

Signal PDFs               & 8.0  & 1.0 & 4.3 & 0.2  & 3.7 & 0.2  \\
Fit bias                 & 2.7  & 1.2 & 0.3 & 0.2  & 2.4 & 0.7  \\
Non-peaking background PDF  & 0.2  & 0.2 & 0.1 & 0.1  & 0.2 & 0.3   \\
BKGIV contribution & 3.9  & 4.0 & 3.2 & 0.7  & 4.9 & 0.5   \\
BKGV contribution         & Negl. & Negl. & Negl. & Negl. & -- & --\\
\hline  
Total                    & 9.3  & 4.3 & 5.4 & 0.8  & 6.6 & 0.9 \\
\hline
\multicolumn{7}{c}{Multiplicative systematic uncertainties ($\%$)}   \\
\hline
Tracking                & 2.0 & 2.0  & 1.0 & 1.0 & 1.5 & 1.5 \\
PID                     & 2.0 & 2.0  & 1.0 & 1.0 & 1.5 & 1.5 \\
$\pi^0$ reconstruction  & 2.0 & 2.0  & 6.0 & 6.0 & 4.0 & 4.0 \\
$\Delta E$ requirement  & 1.7 & 1.7  & 1.7 & 1.7 & 0.3 & 0.3 \\
$K_S^0$ veto             & 0.8 & 0.8 & 1.4 & 1.4 & 0.8 & 0.8  \\ \
$\omega/\eta$ signal region   & 0.2   & 0.2 & 0.2 & 0.2 & 0.2 & 0.2 \\
MC generator      & 2.0 & 3.0 & -- & -- & 3.5 & 3.5 \\
ST yield                & 1.2 & 1.2  & 1.2 & 1.2 & 0.4 & 0.4 \\
Strong-phase in $D^0$ decays   & 7.3  & 0.8 & 7.3  & 7.3  & -- & -- \\
$\mathcal{B}(\omega/\eta \to \pi^+\pi^-\pi^0)$  & 0.8  & 1.2 & 0.8  & 1.2  & 0.8 & 1.2 \\
$\mathcal{B}(\pi^0 \to \gamma \gamma)$  & Negl. & Negl. & Negl. &Negl. &Negl. &Negl.  \\ \hline
Total & 8.7 & 5.3 & 9.9 & 10.0 & 5.9 & 6.0  \\ \hline  \hline
 \end{tabular}
  \label{tab:uncertainties}
  \end{center}
\end{table*}

The uncertainty associated with the DT yield is from the fit procedure
and background subtraction.  The uncertainty from the fit procedure
includes the signal and background shapes as well as the fit bias.  We
perform an alternative 2D fit to the $M_{\rm BC}^{\rm tag}$ versus
$M_{\rm BC}^{\rm sig}$ distribution. The signal $\mathscr{A}$
($\mathscr{B}$) is described with the kernel estimation~\cite{kernel}
of the unbinned 2D (1D) signal MC derived shape convolved with a
Gaussian function. The shape of the background is described with a
modified ARGUS function~\cite{argus} as described above. The relative
changes in the signal yields are taken as the uncertainties.
In this analysis, the 2D fit procedure is validated by repeating the
fit on a large number of pseudo-experiments, which are a mixture of
signals generated with various embedded events and a fixed amount of
background events expected from the real data.  The resultant average
shift of the signal yield is taken as the systematic uncertainty.  As
discussed above, the background BKGIV is estimated with the events in
$\omega / \eta$ sideband regions and incorporating a correction
factor $f$. This induces uncertainties from the definition of sideband
regions and the correction factor.  The uncertainty from sideband
regions is estimated by changing their ranges.  The correction factor
$f$ is determined by fitting the $M_{\ppp}$ distribution of surviving
candidates, which is composed of the events $D\to 5\pi$
including all possible intermediate states ($e.g.$
$\omega/\eta\to\ppp$ or $\rho\to\pi\pi$) and other backgrounds that may
affect $f$.  The procedure to determine $f$ is validated with the inclusive MC sample and
its constituent $D\to 5\pi$ events in the inclusive MC sample.
The resultant $f$ values obtained with these two MC samples
are found to be consistent with each other and data, and the
difference between the two MC results is taken as the uncertainty.
The background BKGV is estimated according to Eq.~\ref{BKGV}, and the
corresponding uncertainties are from the BFs, ST yields and detection
efficiencies, where the first one has been considered as described
above.  Except for the uncertainty related to the $K_S^0$ veto
requirement, which is strongly dependent on the $K_S^0$ mass
resolution, the uncertainties associated with the other requirements
and BFs are fully correlated with those of the signal, and
cancel.  To evaluate the uncertainty associated with the $K_S^0$
veto requirement, we obtain the difference of $K_S^0$ mass resolution
between data and MC simulation using the control sample of $D^0 \to
K_S^0 \pi^+\pi^-\pi^0$.  Then we smear the $M_{\pi\pi}$ distribution
of the background MC samples $D^0\to K_S^0\omega/\eta$ by a Gaussian
function with the differences as parameters. The resultant change of
the efficiency is taken as the uncertainty and is found to be
negligible.


In this analysis, the $\dz\dbar$ pair is from the $\psi(3770)$ decays,
and is quantum correlated, thus additional uncertainty associated with
the strong-phase is considered.
In practice, the absolute BF is calculated as, $\BR^{\rm sig}_{ CP\pm}
= \frac{1}{1-c_f^i(2f_{CP_+}-1)} \BR^{\rm sig}$~\cite{QC}, where
$\BR^{\rm sig}$ is calculated from Eq.~\ref{eq:absBr}, $c_f^i$ are the
strong-phase correction factors of the flavor tags $\bar{D}^0 \to
K^+\pi^-,~K^+\pi^-\pi^0$ and $K^+\pi^-\pi^-\pi^+$~\cite{HFLAV,evans},
and $f_{CP_+}$ is the fraction of the $CP_+$ component of $D^0 \to
\omega/\eta \pi\pi$. The $f_{CP_+}$ value for $D^0 \to \eta
\pi^+\pi^-$ is taken from Ref.~\cite{etatogg}, and the corresponding
systematic uncertainty is determined to be $0.8\%$.  The uncertainties
for $D^0 \to \omega \pi\pi$ and $\eta \pi^0\pi^0$ are 7.3\%, which are
obtained by assuming $f_{CP_+}=0$ or 1 due to the limited
statistics. 
Future BESIII $\psi(3770)$ data will enable a
measurement of the $f_{CP_+}$ of $D^0 \to \omega/\eta \pi^+\pi^-$
decays~\cite{cpc}.


The uncertainties associated with $\BR^{\rm int}$ are obtained from
Ref.~\cite{pdg}.  All the uncertainties discussed above are summarized
in Table~\ref{tab:uncertainties}. The uncertainties associated with
the DT yields, which may affect the significance of observation, are
classified into the additive terms, while the others are multiplicative
terms.  Assuming all the uncertainties to be uncorrelated, the total
uncertainties in the BF measurements are obtained by adding the
individual ones in quadrature.  The $N^{\rm sig}_{\rm DT}$ systematic
uncertainty is given by $\sqrt{{\sigma_{\rm add}}^2 + (\sigma_{\rm
    mult} \times N^{\rm sig}_{\rm DT})^2}$, where $\sigma_{\rm add}$ and
$\sigma_{\rm mult}$ are the total additive and multiplicative
uncertainties, respectively.

\section{Results}
The absolute BFs of $D^0 \to
\omega/\eta \pi^+\pi^-$ and $D^+\to \omega/\eta \pi^+\pi^0$ are
calculated with Eq.~\ref{eq:absBr}.  Since the significance of $D^0
\to \omega/\eta \pi^0\pi^0$ is less than $1\sigma$, we compute
upper limits on the BFs for these two decays at the $90\%$ confidence
level (C.L.) by integrating their likelihood versus BF curves from
zero to $90\%$ of the total curve. The effect of the systematic
uncertainty is incorporated by convolving the likelihood curve with a
Gaussian function with a width equal to the systematic uncertainty.
All results are summarized in Table~\ref{DTYield}.

\section{Summary}
In summary, we perform the BF measurements of SCS decays $D \to \omega
\pi \pi$ using 2.93 $\ifb$ of $\psi (3770)$ data sample collected by
the BESIII detector.  The BFs of $D^0 \to \omega \pi^+\pi^-$ and
$D^+\to \omega \pi^+\pi^0$ are determined to be $(1.33 \pm 0.16 \pm
0.12)\times 10^{-3}$ and $(3.87 \pm 0.83 \pm 0.25)\times 10^{-3}$,
respectively.  The precision of the BF for $D^0 \to \omega
\pi^+\pi^-$ is improved by a factor 2.1 over the CLEO
measurement~\cite{Rubin} and the decay process $D^+\to \omega
\pi^+\pi^0$ is measured for the first time.  These measurements are
important inputs to beauty physics to improve the precision of the CKM
angle $\gamma$ via $B^{\pm} \to D^0 (\to \omega \pi^+\pi^-)
K^{\pm}$~\cite{glw,ggsz} and the semi-tauonic decay $B^0 \to D^{*\pm}
\tau^{\mp}(\to \pi^+\pi^-\pi^{\mp})\nu_{\tau}$~\cite{HFLAV}.  No
evidence of $D^0 \to \omega \pi^0\pi^0$ is found, and the upper limit
on the BF at the $90\%$ C.L. is $1.10\times 10^{-3}$.
Meanwhile, the BFs of $D^0\to\eta \pi^+\pi^-$ and $D^+\to \eta
\pi^+\pi^0$ as well as the upper limit on the BF of $D^0 \to \eta
\pi^0\pi^0$ at $90\%$ C.L.  are measured to be $(1.06 \pm 0.18 \pm
0.07)\times 10^{-3}$, $(2.47 \pm 0.93 \pm 0.16)\times 10^{-3}$, and
less than $2.38\times 10^{-3}$, respectively, with the decay mode
$\eta\to\ppp$. The results are consistent with previous
measurements~\cite{etatogg,pan}.


\section{Acknowledgements}
The BESIII collaboration thanks the staff of BEPCII, the IHEP computing center and the supercomputing center of USTC for their strong support.  This work is supported in part by National Key
Basic Research Program of China under Contract No.
2015CB856700; National Natural Science Foundation
of China (NSFC) under Contracts Nos. 11335008, 11375170, 11425524, 11475164, 11475169, 11605196, 11605198, 11625523, 11635010, 11705192, 11735014, 11822506, 11835012, 11935015, 11935016, 11935018, 11961141012, 11950410506; $64^{\rm th}$ batch of Postdoctoral Science Fund Foundation under contract No. 2018M642516; the Chinese Academy of Sciences (CAS) Large-Scale Scientific Facility Program; Joint Large-Scale Scientific Facility Funds of the NSFC and CAS under Contracts Nos. U1732263, U1832207; CAS Key Research Program of Frontier Sciences under Contracts Nos. QYZDJ-SSW-SLH003, QYZDJ-SSW-SLH040; 100 Talents Program of CAS; INPAC and Shanghai Key Laboratory for Particle Physics and Cosmology; ERC under Contract No. 758462; German Research Foundation DFG under Contracts Nos. Collaborative Research Center CRC 1044, FOR 2359; Istituto Nazionale di Fisica Nucleare, Italy; Ministry of Development of Turkey under Contract No. DPT2006K-120470; National Science and Technology fund; STFC (United Kingdom); Olle Engkvist Foundation under Contract No. 200-0605; The Knut and Alice Wallenberg Foundation (Sweden) under Contract No. 2016.0157; The Royal Society, UK under Contracts Nos. DH140054, DH160214; The Swedish Research Council; U.S. Department of Energy under Contracts Nos. DE-FG02-05ER41374, DE-SC-0012069



\end{document}